\documentclass[10pt, a4paper]{article}

\usepackage[]{lrec-coling2024} 

\title{Echo-chambers and Idea Labs: \\ Communication Styles on Twitter}

\name{Aleksandra Sorokovikova, Michael Becker, Ivan P. Yamshchikov} 

\address{Constructor University, Bremen\\
Technical University of Applied Sciences W{\"u}rzburg-Schweinfurt (THWS),  W{\"u}rzburg \\
CAIRO, Technical University of Applied Sciences W{\"u}rzburg-Schweinfurt (THWS),  W{\"u}rzburg \\
         ivan.yamshchikov@thws.de}

\abstract{
This paper investigates the communication styles and structures of Twitter (X) communities within the vaccination context. While mainstream research primarily focuses on the echo-chamber phenomenon, wherein certain ideas are reinforced and participants are isolated from opposing opinions, this study reveals the presence of diverse communication styles across various communities. In addition to the communities exhibiting echo-chamber behavior, this research uncovers communities with distinct communication patterns. By shedding light on the nuanced nature of communication within social networks, this study emphasizes the significance of understanding the diversity of perspectives within online communities.
 \\ \newline  }

\begin{document}

\maketitleabstract

\section{Introduction}

 Online social environments are often characterised by the phenomenon of the so-called echo chambers  where participants isolate themselves from opposing opinions, reinforcing their own beliefs through limited communication within their community. These chambers can lead to ad hominem attacks, targeting individuals rather than engaging with their arguments, and straw man arguments, which distort opposing viewpoints for easier dismissal. These dynamics contribute to polarization and the adoption of more extreme positions \citep{article1}. Polarization and echo chambers are commonly observed in social networks, facilitated by recommendation algorithms \citep{article2,article3}.

To explore these communities, various algorithms such as the Infomap \cite{rosvall2008maps}, Louvain algorithm \citep{article4}, Stochastic Blockmodels (e.g. \cite{peixoto2020latent}) or force-directed layout \cite{gaisbauer2023grounding} are employed to identify clusters within the user interaction graph. In such setting, users are represented as vertices, and interactions such as social network connections, retweets, and replies on platforms like Twitter (X) are captured as graph edges. These graphs are then partitioned into dense clusters, interpreted as communities sharing similar opinions or engaging in similar activities. Understanding the structure of these communities involves analyzing not only user interactions but also the content they generate. Content analysis allows us to capture the characteristics of the produced content itself \cite{article7}. Previous studies have investigated echo chambers and polarization in social media, particularly concerning topics like COVID-19, proposing models such as Retweet-BERT and DICE for sentiment analysis and detecting ad hominem attacks \cite{article8,article9,article10}.

In this paper, we focus on identifying Twitter (X) communities related to vaccination. We employ community detection algorithms to identify clusters based on user interactions and the content of their tweets. Additionally, we train classifiers for content analysis, such as sentiment, subjectivity, ad hominem, and straw man arguments. Using these classifiers, we evaluate communication style that characterises each community. This approach enables us to uncover that the patterns of user interaction within communities are clearly different. Moreover, to some extent one could identify community of the user based on their communication style. Thus, we suggest that it is sub-optimal to lump every community under a broad umbrella term "echo-chamber". Instead, we suggest there is a need for a more detailed taxonomy based on the detected systematic differences in the communication styles.

\section{Data}

For this work a ready-made twitter (X) vaccination dataset\footnote{\url{https://www.kaggle.com/datasets/keplaxo/twitter-vaccination-dataset}} was taken, it contains approximately 1.5m tweets on the vaccination topic and 770k unique users. This dataset was collected with TWINT - open-source scraping tool. This dataset is suitable for research for several reasons. Firstly, all tweets relate to the topic of vaccination, and the tweets are taken from a very wide time range (starting in 2006, ending in 2019). Secondly, it is possible to build a reply graph, since for tweets that are reply, there is the id of the user to whose tweet this reply was made. A large number of tweets collected in the dataset allows to cover the topic of vaccination from different sides and opinions. Finally, all the discussions are thematically alinged so the differences between texts written by  the members of different communities are less prone to be topically aligned. All the discussions are centered around one general topics and the stylistic differences between the texts are more differentiating than the topics these texts discuss. This makes the dataset a good case-study for the core hypothesis of stylistic distinctions between communications styles of various communities.

  \begin{figure}[hbt!]
      \includegraphics[width=\linewidth]{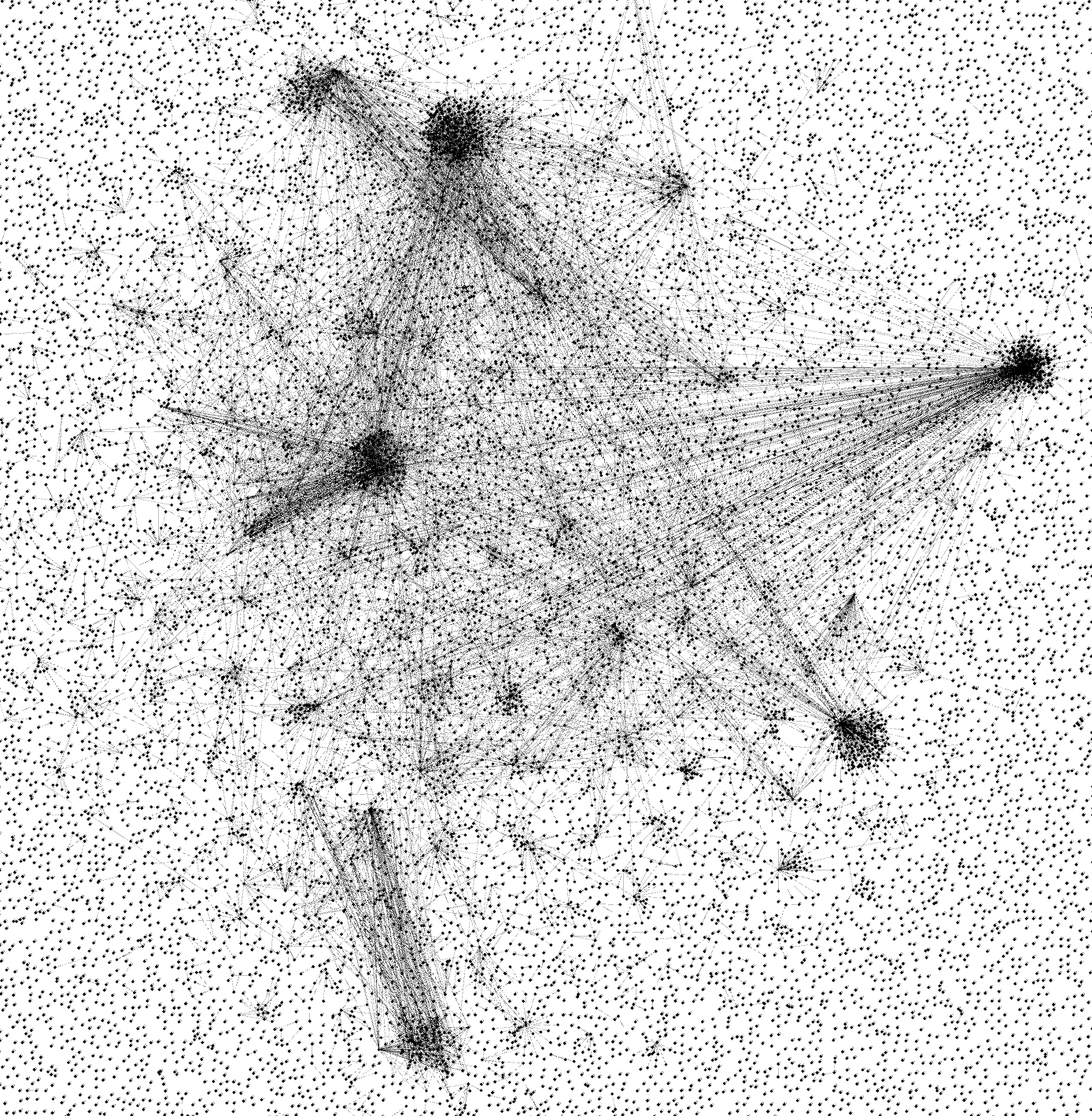}
    \caption{Conversation graph after applying OpenOrd algorithm}
    \label{conversation_graph}
  \end{figure}

  \begin{table*}[ht]
\centering
\begin{tabular}{|c|c|c|c|c|c|c|}
\hline
text/user & 1 & 2 & 3 & 4 &  5 & 6 \\ 
based cluster &  &  &  &  &  & \\\hline
1 & 445** & 942* & 172 & 128 & 70 & 58 \\ \hline
2 & 720** & 2763* & 413 & 323 & 239 & 156 \\ \hline
3 & 262 & 974* & 355** & 71 & 111 & 71 \\ \hline
4 & 237 & 827* & 94 & 264** & 86 & 64 \\ \hline
5 & 202 & 736* & 162 & 82 & 308** & 41 \\ \hline
6 & 205 & 718* & 145 & 92 & 71 & 492** \\ \hline
\end{tabular}
\caption{Confusion matrix for text-based clustering used for user-based cluster prediction. The clusters are unbalanced. Cluster number 2 has the biggest support, thus * markes the highers value in a given row, while ** marks the second highest}
  \label{tab}
\end{table*}

\begin{table}[ht]
\begin{tabular}{|c|c|c|c|c|c|}
\hline
text/user & 1  & 3 & 4 &  5 & 6 \\ 
based cluster   &  &  &  &  & \\\hline
1 & \bf{445}  & 172 & 128 & 70 & 58 \\ \hline
3 & 262  & \bf{355} & 71 & 111 & 71 \\ \hline
4 & 237  & 94 & \bf{264} & 86 & 64 \\ \hline
5 & 202  & 162 & 82 & \bf{308} & 41 \\ \hline
6 & 205  & 145 & 92 & 71 & \bf{492} \\ \hline
\end{tabular}
\caption{Confusion matrix for text-based clustering used for interaction-based cluster prediction. Cluster number 2 is removed. The highest values in a given row are marked bold.}
  \label{tab:2}
\end{table}

\section{Experiments}
We calculate a variety of metrics using texts of the tweets without any information on the user community in which a given tweet is published. Then we cluster the tweets based on those classifiers and compare mutual information between the obtained text-based clustering and the clusters that are formed in the communication graph. It turns out that the textual information alone partly allows to predict the community in which the user communicates.

\subsection{Community clustering} \label{sec:comcl}

Based on the dataset, a graph was built as follows: each vertex represents a user, and an edge is created if user A made a reply to user B. To get rid of noise and make the graph more dense, a weight was added to the edges, which meant the number of replays of the user. Only edges with weight 3 or higher remained in the final graph. Subsequently, the OpenOrd algorithm \citep{article6} was used for spatialization, arranging the vertices in two-dimensional space with $x$ and $y$ coordinates, as depicted in Figure \ref{conversation_graph}. Notably, for community detection, the Louvain algorithm was utilized, with clusters determined based on modularity. The computed modularity value for this graph was 0.902, and the subsequent analysis focused on six of the densest communities, informed by clustering coefficient and modularity considerations.

\subsection{Text Classification Metrics}
We calculate a variety of metrics that assess communication style and are based solely on the textual content of the tweets.

\subsubsection{Polarity Scores}
The polarity scores were calculated using the Optimized BERT Pretraining Approach (RoBERTa), which was trained on around 58 million tweets. We used the pretrained RoBERTa-based classifier developed in \cite{barbieri-etal-2020-tweeteval} to calculate the negativity, neutrality and positivity of a give textual input.

\subsubsection{Subjectivity}
The subjectivity score was determined using the TextBlob \cite{loria2018textblob} library. The subjectivity value in TextBlob indicates the degree of subjective or objective content of a given text. Subjectivity refers to how opinionated or subjective the text is, while objectivity refers to a more factual or objective writing style. The subjectivity value is a floating point value between 0.0 and 1.0, where 0.0 indicates a very objective or factual text and 1.0 indicates a very subjective or opinionated text.



\subsubsection{Logical fallacy}
The two values "label" and "score" were determined using a pre-trained model for logical fallacy detection \citep{article11}. The score label can take values between 0 and 12 and ecodes various logical fallacies, namely: Ad Hominem, Ad Populum, False Causality, Circular Claim, Appeal to Emotion, Fallacy of Relevance, Deductive Fallacy, Intentional Fallacy, Fallacy of Extension, False Dilemma, Fallacy of Credibility, Equivocation. The model assignes a probability between 0 and 1 for every given label.

\section{Predicting Community Membership with Communication Style}

Now every tweet could be described by a set of various classifier scores. At the same time we know the community to which the author of the tweet belongs since we have the structure of the reply clusters that we obtained in Section \ref{sec:comcl}. To test whether every echo-chamber is characterised with similar communication style we can build clusters of tweets based solely on the classifier scores. Normally, a choice for the number of clusters in a clustering could be difficult. However, since the reply graph clustering procedure has already detected six clusters we can choose six clusters for our text-based clustering as well.  Now every tweet belongs to one text-based cluster as well as one reply graph based cluster. Comparing those labels allows us to see to which extent solely textual content of the tweet informs us on the community in which given communication occurs.

\begin{figure}[hbt!]
      \includegraphics[width=\linewidth]{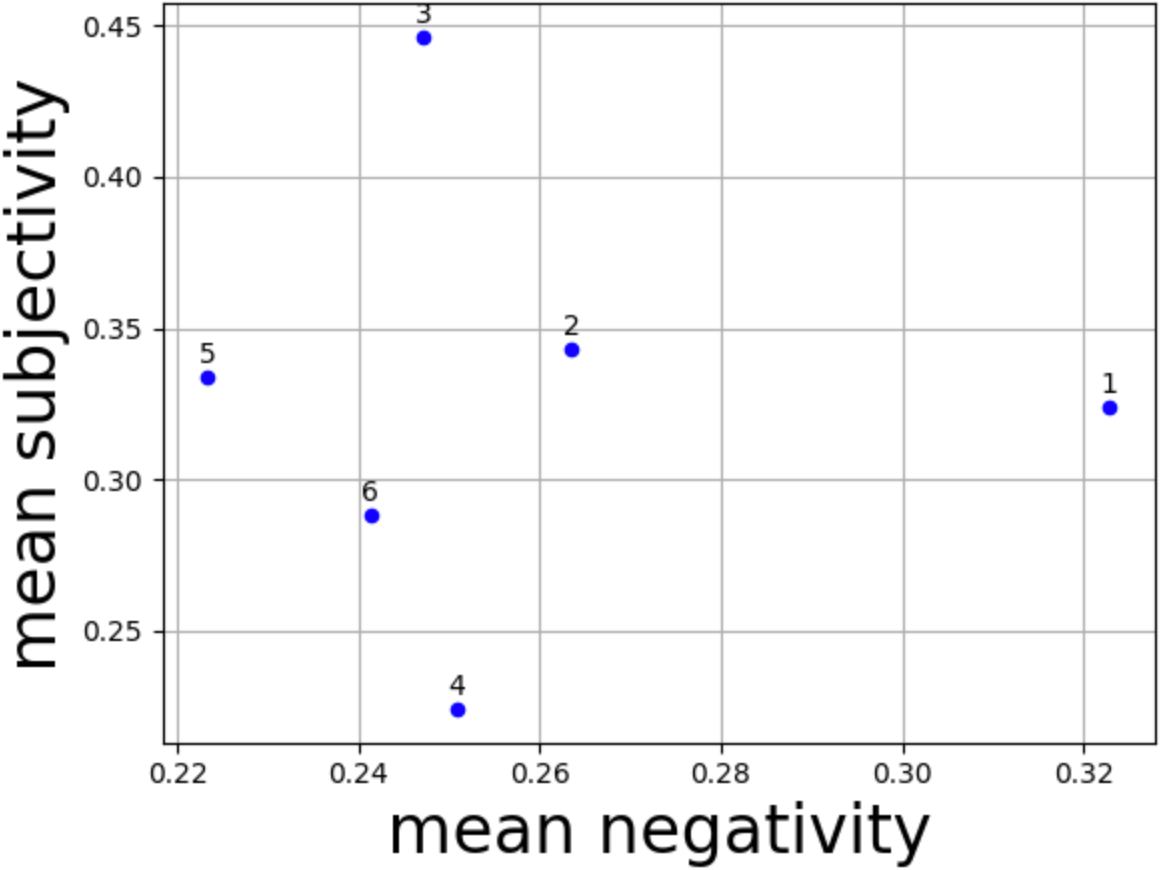}
    \caption{Average scores for mean subjectivity and mean negativity in all 6 user communities. One can clearly see that one of the community is characterised as highly subjective while another is highly negative while scoring lower on subjectivity.}
    \label{cluster_differences}
  \end{figure}

Figure \ref{cluster_differences} demonstrates average scores for six clusters detected in Section \ref{sec:comcl}. Once immediately sees that all six are characterize by rather different communication styles. One can see that cluster number two is represented by the point in the "center". It is the biggest cluster that includes a lot of weakly connected users that do not form a dense clique. This cluster has the biggest support in terms of absolute numbers but represents users who are not active members of any on of the five dense communities but are rather occasional posters. Thus, it stands to reason that the average scores of the classifiers for the tweets in this cluster end up in the center of the cloud of points. Table \ref{tab} shows the confusion matrix between knn-clusters based on texts of the tweets and the clusters obtained from the graph of interactions between users. The accuracy of the user-based clustering when predicted by text is 0.35 which is quite interesting in itself.

 If we drop this cluster number two that represents weakly connected users that do not form a tight community the accuracy reaches 0.44, see Table \ref{tab:2}. This highlights that tight communities have  have a distinct expressive communication style indeed.

 \section{Discussion}
This study offers an case study and an initial exploration of the varied communication styles within Twitter (X) communities discussing vaccination, challenging the simplistic notion of echo chambers. Through the application of community detection algorithms and text classification metrics, we demonstrate significant diversity in subjectivity, negativity, and logical fallacies across different groups. This suggests a broader spectrum of communicative behaviors in online discussions than typically discussed.

Our findings highlight the importance of nuanced understanding of online discourse, pointing towards the necessity for a more detailed taxonomy of communication styles. This work suggests a pathway for future research to explore the complexities of digital community communication, advocating for a deeper examination beyond conventional categorizations to better understand how these communities interact and evolve.

\section{Conclusion}

In our investigation of communication patterns within Twitter's (X's) echo chambers, we aimed to identify variations in discourse, specifically seeking environments akin to 'Idea Labs' where open, critical discussion prevails over personal attacks. Utilizing computational methods, we analyzed a substantial dataset to discern these communication styles.

Our results did not confirm the presence of 'Idea Labs' in the studied dataset. However, the study revealed significant variations in communication styles across different echo chambers. Despite discussing identical topics, the textual characteristics within each community were distinct enough to allow for a predictive model to accurately categorize tweets based on their origin.

This finding is critical, demonstrating the extent to which echo chambers can influence discourse style. It also highlights the potential for computational approaches to identify and categorize such patterns in online communication.

\section*{Limitations}

This paper focuses on one particular case-study. The topic of the discussions is specific and all the results are limited to twitter  (X) discussions only. Thus one can not be sure that the results are general and could be applicable to other social media communities or to the discussions around other topics. However, the provided case-study is a good starting point to initiate a deeper discussion of a more nuances approach to community formation in social media that regards the phenomenon in a more holistic manner and takes into account both the structure of the social graph as well as the content of the communications.

\section*{Acknowledgements}

The authors are deeply thankful to Dr. Felix Glaisbauer for his support, fruitful discussions and valuable advice.

\appendix




\nocite{*}
\section{Bibliographical References}\label{sec:reference}

\bibliographystyle{lrec-coling2024-natbib}
\bibliography{lrec-coling2024-example}


\end{document}